\documentclass[aps,prl,twocolumn,reprint,showkeys,showpacs,longbibliography]{revtex4-1}
\usepackage{amsmath}
\usepackage{amssymb}
\usepackage{graphicx}
\usepackage{hyperref}
\usepackage{color,xcolor}
\usepackage{hyperref}
\begin{document}
\title{Orientation-dependent ferroelectricity of strained PbTiO$_3$ films}
\author{Huimin Zhang}
\author{Ming An}
\author{Xiaoyan Yao}
\email{yaoxy@seu.edu.cn}
\author{Shuai Dong}
\email{sdong@seu.edu.cn}
\affiliation{Department of Physics, Southeast University, Nanjing 211189, China}
\date{\today}

\begin{abstract}
PbTiO$_3$ is a simple but very important ferroelectric oxide that has been extensively studied and widely used in various technological applications. However, most previous studies and applications were based on the bulk material or the conventional [$001$]-orientated films. There are few studies on PbTiO$_3$ films grown along other crystalline axes. In this study, a first-principles calculation was performed to compute the polarization of PbTiO$_3$ films strained by SrTiO$_3$ and LaAlO$_3$ substrates. Our results show that the polarization of PbTiO$_3$ films strongly depends on the growth orientation as well as the monoclinic angles. Further, it is suggested that the ferroelectricity of PbTiO$_3$ mainly depends on the tetragonality of the lattice, instead of the simple strain.
\end{abstract}


\maketitle

\section{Introduction}
Heterostructures and thin films of transition metal oxides are not only physically interesting but also technologically important, and have become a rapidly developing branch of condensed matter physics and materials science \cite{Dagotto:Sci07,Mannhart:Sci,Hwang:Nm,Choi:Nc,Jiang:Nl}. In heterostructures of correlated oxide films, charge, spin, and lattice can be reconstructed at the interfaces, generating emergent phenomena such as two-dimensional electron gas, superconductivity, new magnetic orders, and improper ferroelectricity \cite{Dong:Prb13,Zhang:PRB15,Duan:Fp}. The fabrication of thin films involves many tunable parameters, such as parent materials, strain from substrate, thickness of films, multi-layer configurations, and stoichiometric control.

In addition, the growth orientation is a crucial condition in tuning the physical properties of films and heterostructures. For example, in LaFeO$_3$-LaCrO$_3$ superlattices, ferromagnetism appears when Fe and Cr layers are atomically stacked along the pseudocubic [$111$] direction, whereas antiferromagnetism appears in the cases of superlattices growing along the [$001$] and [$110$] directions \cite{Ueda:Jap,Zhu:Jap}. Furthermore, in LaNiO$_3$-LaMnO$_3$ superlattices, the induced magnetization is largest in the ($111$)-stacking direction but weakest in the ($001$)-stacking direction \cite{Gibert:Nm,Dong:Prb13}.

Even for a single material film, the growth orientation may alter the ground state. In a recent study, we found that the YTiO$_3$ film on LaAlO$_3$ becomes A-type antiferromagnetic when growing along the [$001$]-axis \cite{Huang:Jap}, while it retains ferromagnetism when growing along the [$110$]-axis; this finding agrees with experimental observations \cite{Huang:Jap2,Chae:Apl}. For LaTiO$_3$, the exchange coefficients and, thus, the Neel temperature were predicted to be significantly enhanced in the [$111$] bilayer compared with the [$001$] one \cite{Weng:Jap2}. In addition, for SrMnO$_3$ films grown on SrTiO$_3$ ($111$) substrates, a hexagonal ($4$H) structure is more stable than the cubic phase \cite{Song:Fp}.

Despite the abundant researches on magnetic systems, the orientation-dependence of ferroelectricity has rarely been studied. PbTiO$_3$ is one of the simplest and most important ferroelectric oxides, and hence, it is selected as the model system for demonstrating the physical mechanism in the present work. It clearly exhibits single transition at $T_C=763-766$ K \cite{Sani:Jssc,Lee:APL} from a paraelectric phase with a cubic structure to a ferroelectric phase. At room temperature, PbTiO$_3$ has a tetragonal structure (space group $P4mm$), and the experimental measured lattice constants are: $a=3.904$ {\AA}, $c=4.158$ {\AA}, giving it a moderate tetragonality {($c$/$a$=$1.065$)}. Both Ti$^{4+}$-oxygen octahedra and Pb$^{2+}$ contribute to the spontaneous polarization, which is as large as $\sim75-80$ $\mu$C/cm$^2$ \cite{Gavrilyachenko:Spss,Haun:Jap,Sharma:Prb}. The PbTiO$_3$ bulk and [$001$]-orientated films have been extensively studied and widely used in applications \cite{Bousquet:Nat,Tang:Sci}. However, there are very few reports on PbTiO$_3$ films with other orientations such as the [$110$] and [$111$] directions, which are physical nontrivial as well as technologically important.

In this work, we use first-principles calculations to study the epitaxial strain effects on the ferroelectricity of PbTiO$_3$ films grown along different orientations.
Our study finds that the polarizations of the different orientations are considerably different. The ferroelectricity of PbTiO$_3$ film is mainly governed by the tetragonality, instead of the simple strain.

\begin{figure*}
\centering
\includegraphics[width=0.3\textwidth]{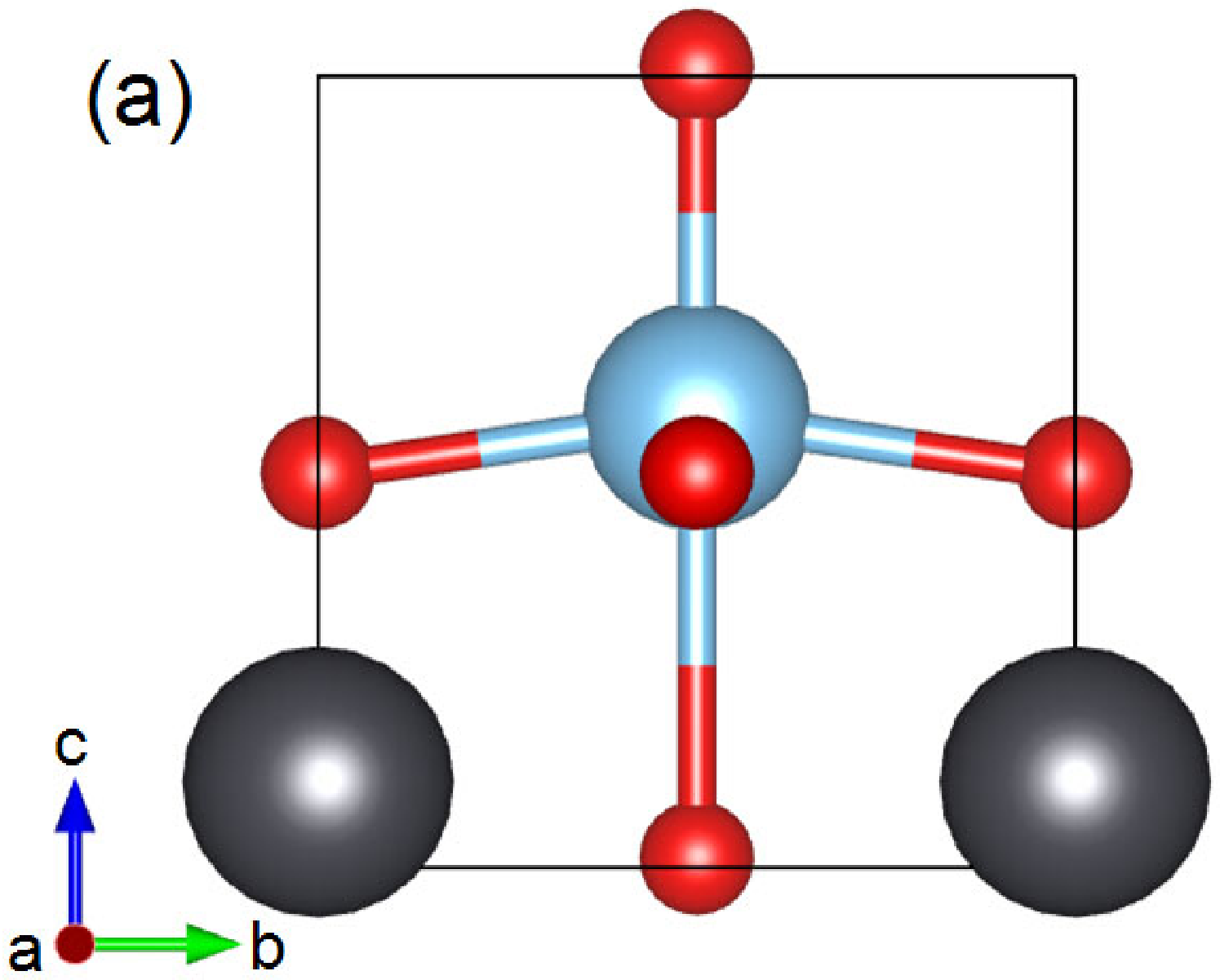}\includegraphics[width=0.7\textwidth]{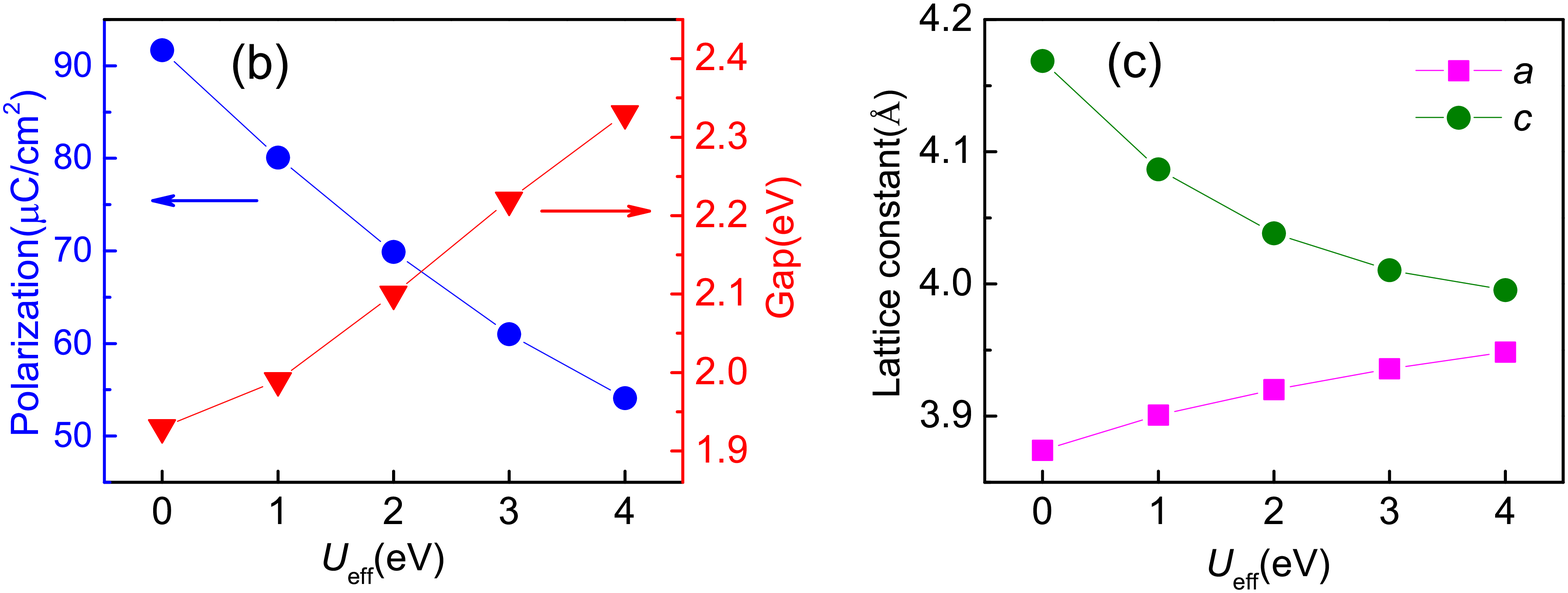}
\caption{(a) The structure of tetragonal PbTiO$_3$. (b-c) Physical properties of PbTiO$_3$ in our DFT calculation as a function of $U$$_{\rm eff}$. (b) The blue line is polarization and the red line is band gap. (c) The lattice constants.}
\label{bulk}
\end{figure*}

\section{Model \& method}
Cubic SrTiO$_3$(STO), cubic LaAlO$_3$(LAO), and rhombohedral LaAlO$_3$ substrates were adopted for comparison. The in-plane lattice constants were fixed to the lattice constants of these substrates to simulate the strain, even though these substrates do not really appear in the calculations. Three growth orientations, [$001$], [$110$], and [$111$], were also compared.

The first-principles theory calculations were based on the projected augmented wave pseudopotentials using the Vienna ab initio simulation package (VASP) \cite{Kresse:Prb,Kresse:Prb96}. The electron interactions are described using PBEsol (Perdew--Burke--Ernzerhof--revised) \cite{Perdew:Prl08} parametrization of the generalized gradient approximation plus $U$ (GGA+$U$) method \cite{Perdew:Prl, Kresse:Prb99, Blochl:Prb2}. The choice of PBEsol is preferable for better description of the lattice structure of titanates, while the traditional PBE often fails for ferroelectric titanates. The Dudarev implementation \cite{Dudarev:Prb} is adopted with an on-site Coulomb interaction $U$$_{\rm eff}=U-J$ applied to the $3$d orbitals of Ti. The atomic positions are fully optimized iteratively until the Hellman--Feynman forces converged to less than $0.01$ eV/{\AA}. The plane-wave cutoff is set to $500$ eV. The Monkhorst--Pack $k$-point meshes are $7\times7\times1$, $5\times7\times7$, $5\times7\times5$ and $5\times3\times4$ for the [$001$]-, [$110$]-, [$011$]- and [$111$]-oriented films, respectively. These crystalline axes denote a pseudocubic one. In the following sections, the $a$-$b$ axes denote the in-plane coordinates while the $c$-axis denotes the out-of-plane one. The Berry phase method is adopted to calculate the ferroelectric polarization \cite{Resta:Rmp}.

\section{Results \& discussion}
\subsection{PbTiO$_3$ bulk}
First, the physical properties of unconstrained PbTiO$_3$ were investigated. The tetragonal structure of the ferroelectric state was fully optimized with various $U_{\rm eff}$ from $0$ eV to $4$ eV in steps of $1$ eV. Then the fully relaxed structure was used to calculate the polarization and band gap. As shown in Fig. 1.(b), the polarization decreases from $91.7$ $\mu$C/cm$^2$ to $54.1$ $\mu$C/cm$^2$ and the band gap increases from $1.9$ eV to $2.3$ eV with increasing $U_{\rm eff}$.

When $U_{\rm eff}$=$1$ eV, the fully relaxed lattice gives: $a=3.901$ {\AA}, $c=4.087$ {\AA} and $c/a=1.047$, which are close to the experimental values. Further, the polarization is just $80.1$ $\mu$C/cm$^2$, which is close to the previous theoretical value $88$ $\mu$C/cm$^2$ \cite{Gotthard:Prl} and experimental value $\sim75-80$ $\mu$C/cm$^2$ \cite{Gavrilyachenko:Spss,Haun:Jap,Sharma:Prb}. These results imply that $U_{\rm eff}$=1 eV is a suitable coefficient for GGA+$U$ calculation based on the PBEsol exchange to describe PbTiO$_3$. Although the calculated band gap of $\sim2.0$ eV is lower than the experimental value ($\sim3.4$ eV) \cite{Piskunov:Cms}, the result remains acceptable considering the well-known underestimation of band gap in DFT calculations. Therefore, $U_{\rm eff}=1$ eV will be chosen as the default coefficient in the following calculations.

\subsection{PbTiO$_3$ along the [$001$] axis}
Next, calculations were performed for PbTiO$_3$ films with epitaxial strains. The in-plane lattice constants were fixed to match the particular surface of a substrate. Then, the out-of-plane lattice constant ($c$) and internal atomic positions were fully optimized. More precisely, the value of $c$ was obtained by searching for the lowest energy. The most conventional [$001$]-orientated films were studied first.

The in-plane lattice constant of cubic STO is $3.905$ {\AA}, which is very close to that of PbTiO$_3$ itself. Thus, the tensile strain is almost negligible. The relaxed lattice constant along the $c$-axis is $4.088$ {\AA}. The polarization was calculated to be $80.4$ $\mu$C/cm$^2$  along the $c$-axis, which is very close to the value of bulk PbTiO$_3$. For comparison, another substrate with a smaller lattice, LaAlO$_3$, is tested. The cubic lattice constant of LaAlO$_3$ is about $3.8106$ {\AA}($821$ K) \cite{Howard:Jpcm}, which can give a strong compressive strain to the PbTiO$_3$ film. As expected, the polarization was $84.7$ $\mu$C/cm$^2$.

In fact, at room temperature, LAO is rhombohedral (space group $R\bar{3}c$, No. $167$) instead of cubic. Here, we also calculated the more realistic condition with a rhombohedral substrate (still using its pseudocubic ($001$) surface). The unit cell was rotated to ensure that the PbTiO$_3$ epitaxial grown on rhombohedral LAO can be applied as the substrates in our calculation. The $a$ and $b$ axes of PbTiO$_3$ were fixed to the rhombohedral LAO substrate. Further, owing to the strong compressive strain, an even larger polarization of $105.3$ $\mu$C/cm$^2$ was obtained, which was also along the $c$-axis.

\begin{figure}
\centering
\includegraphics[width=0.48\textwidth]{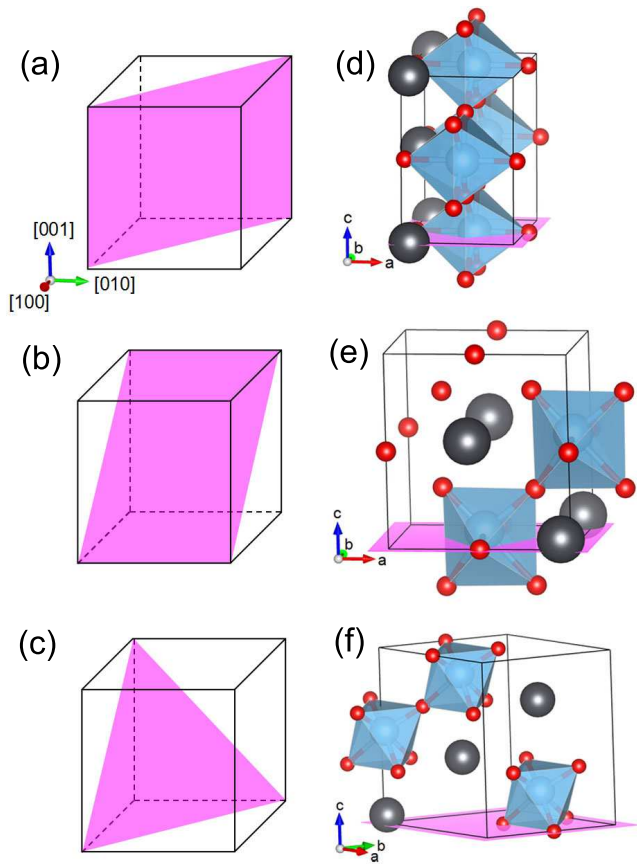}
\caption{(a-c) The pink planes represent ($110$), ($101$) and ($111$) planes respectively. (d) The PbTiO$_3$ film in [$110$] direction. (e) The PbTiO$_3$ film in [$101$] direction. (f) The PbTiO$_3$ film in [$111$] direction.}
\label{superlattice
}
\end{figure}

\subsection{PbTiO$_3$ on the ($110$)-faced substrates}
The in-plane lattice constants $a$ and $b$ of the ($110$)-faced STO substrate are $3.905$ {\AA} and $5.5225$ {\AA} ($=3.905{\times}\sqrt{2}$), respectively. For this substrate, two growing directions are possible for the PbTiO$_3$ film. One is along the [$110$] axis of PbTiO$_3$, shown in Fig.2(d), with the polarization in plane. Another is PbTiO$_3$ grown along the [$101$] axis or its equivalent axes, shown in Fig.2(e), with both in-plane and out-of-plane polarization. Both cases have been investigated.

For the first case, the optimal out-of-plane lattice constant is $5.59$ {\AA} (the [$110$] axis of pseudocubic coordinate). The polarization of this strained PbTiO$_3$ film is only $59.3$ $\mu$C/cm$^2$, which is much weaker than the original polarization. This is reasonable since the polar direction along the [$001$] axis was clamped by the substrate (the $a$-axis).

This scenario can be easily confirmed by using the LAO substrate. For the cubic LAO substrate, the smaller in-plane lattice constants $a=3.8106$ {\AA} and $b= 5.389$ {\AA} generate an even smaller polarization $35.5$ $\mu$C/cm$^2$ since the polar direction was considerably more compressed. When growing on the rhombohedral LAO substrate ($a= 3.791$ {\AA} and $b= 5.3655$ {\AA}), the polarization of the strained PbTiO$_3$ was further suppressed down to $24.7$ $\mu$C/cm$^2$. Considering the direction of polarization, the tetragonality should be redefined as $a/c$, instead of $c/a$. Thus, the smaller in-plane lattice constant $a$ of the substrate would lead to a smaller polarization.

In contrast, for the second case, namely the [$101$]-axis-growing film, the results are quite different, as shown in Fig.2(e). The polarization of the PbTiO$_3$ films was nearly identical to its intrinsic value despite the substrates, even if the lattice constants of the strained PbTiO$_3$ on the three substrates were different. Although the in-plane component of polarization was compressed, as in the above [$110$]-axis case, the out-of-plane component was enhanced significantly, characterized by the displacement of large ions along the $c$-axis.
\begin{table}
\caption{The ferroelectric polarization and relaxed lattice constants of PbTiO$_3$ films strained by different substrates. C-LAO is the cubic LaAlO$_3$ substrate. R-LAO is the rhombohedral LaAlO$_3$ substrate. For the [$111$] orientated film on STO, two values are presented: the larger one is obtained for the lattice with monoclinic distortion, while the smaller one in parenthesis is for the lattice with $c\bot ab$. }
\centering
\begin{tabular*}{0.48\textwidth}{@{\extracolsep{\fill}}lllll}
\hline \hline
\centering
[$001$]&substrate & STO & C-LAO & R-LAO\\
\hline
 & $a$({\AA}) & $3.905$ & $3.8106$ & $5.3655$\\
    & $b$({\AA}) & $3.905$ & $3.8106$ & $5.357$\\
    & $c$({\AA}) & $4.0885$ & $4.154$ & $4.36$\\
    & $P$($\mu$C/cm$^2$) & $80.4$ & $84.7$ & $105.3$ \\

    \hline \hline
[$110$]&substrate & STO & C-LAO & R-LAO\\
\hline
  & $a$({\AA}) & $3.905$ &$3.8106$ & $7.582$\\
    & $b$({\AA}) & $5.5225$ & $5.389$ & $5.3655$\\
    & $c$({\AA}) & $5.59$ & $5.67$ & $5.71$\\
    &$P$($\mu$C/cm$^2$) & $59.3$ & $35.5$ & $24.7$ \\

\hline \hline
[$101$]&substrate & STO & C-LAO & R-LAO\\
\hline
  & $a$({\AA}) & $5225$ & $5.389$ & $5.3655$\\
    & $b$({\AA}) & $3.905$ & $3.8106$ & $7.582$\\
    & $c$({\AA}) & $5.72$ & $5.79$ & $5.81$\\
    &$P$($\mu$C/cm$^2$) & $77.5$ & $76.8$ & $79.4$ \\

\hline \hline
[$111$]&substrate & STO & C-LAO & R-LAO \\
\hline
 &$a$({\AA}) & $5.5225$ & $5.389$ & $5.3655$\\
&$b$({\AA}) & $5.5225$ & $5.389$ & $5.3655$\\
&$c$({\AA}) & $6.91$ & $7.06$ & $7.08$\\
&$P$($\mu$C/cm$^2$) & $82.4$ ($64.7$) & $63.0$ & $62.6$ \\
\hline \hline
\end{tabular*}
\label{table2}
\end{table}

Then, which case will be the most possible one in real thin films. The calculated energy of the [$101$]-oriented films were lower than the [$110$]-oriented films by $26$ meV, $65$ meV and $72$ meV per Ti for cubic STO, cubic LAO, and rhombohedral LAO substrates, respectively. In this sense, when using the ($110$)-faced substrates, the strained PbTiO$_3$ will keep a large polarization (similar to the [$101$] case in Table (I)), whose direction deviates away from the original one but is still mostly out of plane.

\subsection{PbTiO$_3$ on the ($111$)-faced substrates}
According to a previous DFT study on [$111$] epitaxially strained BaTiO$_3$ and PbTiO$_3$ \cite{Oja:Prb08}, the total polarizations of the stable phases of PbTiO$_3$ were observed to be almost independent of the applied strain, while the application of compressive strain suppressed the polarization of BaTiO$_3$. However, the study was performed on a reduced unit cell containing only $5$ atoms, which may prohibit some modes of lattice distortions. Thus, a systematic study remains necessary.

The schematic structure of the [$111$]-orientated film is shown in Fig.2(f). The minimum unit cell contains $15$ atoms. First, the normal relaxation process is used, with the in-plane ($a$- and $b$-axes) lattice constants fixed and out-of-plane lattice ($c$-axis) optimized. Here the $c$-axis is perpendicular to the $a$-$b$ plane. In this case, the ferroelectric polarizations in all PbTiO$_3$ films are suppressed, as listed in Table I. The reason is similar to that for the aforementioned [$110$]-oriented film, namely the in-plane clamping is a disadvantage to the tetragonality along the [$001$]-axis.

We have also tested the monoclinic distortion, by allowing $c$-axis deviation from the perpendicular direction of the $a-b$ plane. For the ($111$) STO substrate, a deviation is preferred, giving $89.0$$^{\circ}$ as the most stable state. This tiny monoclinic distortion can restore the tetragonality of PbTiO$_3$, resulting in a large polarization of $82.4$ $\mu$C/cm$^2$. This result agrees with that of the previous study.

However, for the two-type LAO substrates, the optimized $\alpha=\beta=90 ^{\circ}$ suggests that monoclinic distortion will not occur. In other words, the polarization will be greatly reduced on the ($111$)-faced LAO substrates. It should be noted that the strain is beyond the region covered in the previous study. In short, the ferroelectric properties of [$111$]-oriented PbTiO$_3$ films depend on both the strain as well as the monoclinic distortion.

Finally, it should be noted that although only PbTiO$_3$ was studied here as a model system, the physical mechanism revealed in this work should by generally valid for other perovskite ferroelectrics such as BaTiO$_3$. The origin of ferroelectricity in PbTiO$_3$ is two-fold, arising from both Pb$^{2+}$ and Ti$^{4+}$, which leads to a very large polarization. In BaTiO$_3$, the ferroelectric is much weaker, without a large contribution from Pb$^{2+}$. Otherwise, their ferroelectric physics is very similar. Therefore, their behavior will be quite similar upon strain, even though quantitatively their lattice constants (and thus the strain intensity) are somewhat different.

\section{Conclusion}
In summary, the ferroelectric polarizations of PbTiO$_3$ films grown on the ($001$)-, ($110$)- and ($111$)-faced STO and LAO substrates have been studied in detail. The polarization of the [$001$]-oriented PbTiO$_3$ film on STO is very close to that of the bulk material, since their lattice constants match very well, whereas the polarization on LAO substrates becomes a little larger than that of the bulk owing to in-plane compression. For the [$110$]- and [$101$]-oriented PbTiO$_3$ films, the in-plane ferroelectric component was significantly reduced owing to the suppressed tetragonality, the latter of which can persist a large out-of-plane ferroelectric polarization. In the [$111$]-oriented film, the polarization will be either largely suppressed or restored depending on the substrates (LAO vs. STO). On LAO, the tetragonality is suppressed, whereas on STO, the spontaneous monoclinic distortion helps to restore the tetragonality.

\acknowledgments{Work was supported by the National Natural Science Foundation of China (Grant Nos. 51322206 and 11274060), the Natural Science Foundation of Jiangsu Province of China (Grant No. BK20141329).}

\bibliography{ref}
\end{document}